\documentclass[
  a4paper, 
  12pt, onecolumn,
 ]{article}

\title{\bf     Bethe Ansatz calculation of   
               the spectral gap of the asymmetric exclusion process  
}
\author{        O. Golinelli, K. Mallick
\bigskip
\\ \ad          email: \{golinelli, mallick\}@cea.fr
\\ \ad          Service de Physique Th\'eorique, 
\\ \ad          Cea Saclay, 91191 Gif-sur-Yvette, France
}
\date{\normalsize       December 15, 2003
\\                      Preprint T03/196 ; arXiv:cond-mat/0312371
}

\newcommand  {\ad}{\normalsize\em}      
\usepackage[final]{graphicx}
\newcommand{\figwidth}{\columnwidth}

\newcommand{\Li}{\mathrm{Li}}   
\newcommand{\Arg}{\mathrm{Arg}}

\begin{document}
\maketitle

\begin{abstract}
\normalsize

    We present a new derivation of the spectral gap of the totally
 asymmetric exclusion process on a half-filled ring of size $L$ by using
 the Bethe Ansatz. We show that, in the large $L$ limit, the Bethe
 equations reduce to a simple transcendental equation involving the
 polylogarithm, a classical special function.  By solving that equation,
 the gap and the dynamical exponent are readily obtained.  Our method can
 be extended to a system with an arbitrary density of particles.


\medskip \noindent Pacs numbers: 05.40.-a; 05.60.-k.

\end{abstract}

\section{Introduction}

The asymmetric simple exclusion process (ASEP) is a model of driven
diffusive particles on a lattice with hard-core exclusion (for general
review see Spohn 1991).  The ASEP appears as a minimal building block in a
large variety of models for hopping conductivity (Richards 1977), polymer
reptation (Widom {\it et al.} 1991), traffic flow (Schreckenberg and Wolf
1998) or surface growth (Krug 1997). In particular, the ASEP in one
dimension is a discrete version of the Kardar-Parisi-Zhang (KPZ) equation
(Halpin-Healy and Zhang 1995).  In biophysics, the ASEP has been used to
describe the diffusion of macromolecules through narrow vessels (Levitt
1973) and the kinetics of protein synthesis on RNA (MacDonald and Gibbs
1969); more recently, a mapping between sequence-alignment and the
exclusion process has been proposed (Bundschuh 2002). From a theoretical
point of view, the ASEP plays the role of a paradigm in non-equilibrium
statistical mechanics: it displays a variety of features such as boundary
induced phase transitions (Krug 1991), spontaneous symmetry breaking in
one dimension (Evans {\it et al.} 1995) and dynamical phase separation
(Evans {\it et al.} 1998).

Exact results for the ASEP in one dimension have been derived using two
complementary approaches (for a review see Derrida 1998).  The Matrix
Ansatz (Derrida {\it et al.} 1993) allows to calculate steady state
properties such as invariant measures (Speer 1993), current fluctuations in
the stationary state and large deviation functionals (Derrida {\it et al.}
2003). The Bethe Ansatz (Dhar 1987) provides spectral information about the
evolution (Markov) operator (Gwa and Spohn 1992; Sch\"utz 1993; Kim 1995)
which can be used to derive large deviation functions (Derrida and Lebowitz
1998; Derrida and Appert 1999; Derrida and Evans 1999).  The exact relation
between these two techniques is still a matter of investigation (Alcaraz
{\it et al.} 1994; Stinchcombe and Sch\"utz 1995; Alcaraz and Lazo 2003).

The relaxation time to the stationary state for the ASEP on a lattice of
size $L$ scales typically as a power law, $L^z$, $z$ being the ASEP
dynamical exponent.  The calculation of $z$ for a one dimensional system by
Bethe Ansatz is an important exact result that was first announced by Dhar,
who found $z = 3/2$ (Dhar 1987).  The spectral gap (and thus $z$) was
subsequently calculated for the half-filling case by Gwa and Spohn (1992)
and for an arbitrary density by Kim (1995) who mapped the ASEP into a
non-Hermitian XXZ Heisenberg spin chain.  The one-dimensional ASEP belongs
to the KPZ universality class and therefore the KPZ dynamical exponent in
one dimension is equal to 3/2; this result was previously deduced from
Galilean invariance and renormalization group arguments (for a review see
Krug 1997).

In this work, we present a new method of calculating the spectral gap of
the totally asymmetric exclusion process (TASEP) starting from the Bethe
Ansatz equations. Our method, based on an analytic continuation formula,
circumvents the technical difficulties involved in the derivation of Gwa
and Spohn (1992) and renders the calculation of the ASEP dynamical exponent
much more concise and transparent.  Besides, our technique can be readily
extended to the arbitrary density case and allows us to calculate the
spectral gap for the asymmetric exclusion process with a tagged particle.

The outline of this paper is as follows. In section 2, we recall the
definition of the TASEP, present the Bethe Ansatz equations without
deriving them and summarize their analysis.  In section 3, we present our
original calculation of the spectral gap in the half-filling case.
Concluding remarks and generalizations of our method are given in section~4.

\section{Bethe  Ansatz  for the TASEP}

\subsection{The TASEP  model} 

We consider the totally asymmetric simple exclusion process on a periodic
one dimensional lattice with $L$ sites (sites $i$ and $L + i$ are
identical). In this model, the total number $n$ of particles is conserved.
Each lattice site $i$ ($1 \le i \le L$) is either occupied by one particle
or is empty ({\em exclusion rule}). Stochastic dynamical rules govern the
evolution of the system: a particle on a site $i$ at time $t$ jumps, in
the interval between times $t$ and $t+dt$, with probability $dt$ to the
neighbouring site $i+1$ if this site is empty. The total number of
configurations for $n$ particles on a ring with $L$ sites is given by
$\Omega = L! / [ n! (L-n)!]$.  In the stationary state, all configurations
have the same probability $1/\Omega$ (Derrida 1998).

A configuration can be characterized by the positions of the $n$ particles
on the ring, $(x_1, x_2, \dots, x_n)$ with $1 \le x_1 < x_2 < \dots < x_n
\le L$. We call $\psi_t(x_1,\dots, x_n)$ the probability of this
configuration at time $t$; the probability distribution $\psi_t$ of the
system at time $t$ is thus a $\Omega$-dimensional vector.  As the ASEP is
a continuous-time Markov ({\it i.e.}, memoryless) process, the time
evolution of $\psi_t$ is determined by the master equation
\begin{equation}
    \frac{d\psi_t}{dt} = M \psi_t  \, , 
\end{equation}
where the transition rate $\Omega\times\Omega$ matrix $M$ is the Markov
matrix.  A right eigenvector $\psi$ is associated with the eigenvalue $E$
of $M$ if
\begin{equation}
  M \psi = E\psi      \, . 
  \label{eq:mpsi=epsi}
\end{equation}
The Markov matrix $M$ is a real non-symmetric matrix and, therefore, its
eigenvalues (and eigenvectors) are either real numbers or complex
conjugate pairs. The spectrum of $M$ contains the eigenvalue $E = 0$ and
the associated right eigenvector is the stationary state $\psi(x_1,\dots,
x_n) = 1/\Omega$.  Because the dynamics is ergodic ({\it i.e.}, $M$ is an
irreducible and aperiodic Markov matrix), the Perron-Frobenius theorem
implies that all eigenvalues $E$ except $0$ have a strictly negative real
part; the relaxation time of the corresponding eigenmode is $\tau =
-1/\mathrm{Re}(E)$.  (The imaginary part of $E$ gives rise to an oscillatory
behaviour).

In this paper, we shall calculate the gap $E_1$, {\it i.e.}, the non-zero
eigenvalue of $M$ with largest real part. The eigenmode associated with
$E_1$ has thus the longest relaxation time that scales as $L^z$, $z$
being the dynamical exponent.

\subsection{The Bethe equations}

Writing $M$ explicitly, the eigenvalue equation~(\ref{eq:mpsi=epsi})
becomes
\begin{equation}
  E \psi(x_1,\dots, x_n) = \sum_i \left[ 
              \psi(x_1, \dots, x_{i-1},\ x_i-1,\ x_{i+1}, \dots, x_n) 
              - \psi(x_1,\dots, x_n) \right] \, , 
\end{equation}
where the sum runs over the indexes $i$ such that $ x_{i-1} < x_i-1$, {\it
i.e.}, such that the corresponding jump is allowed.  The {\em Bethe Ansatz}
assumes that the eigenvectors $\psi$ can be written in the form
\begin{equation}
  \psi(x_1,\dots,x_n) = \sum_{\sigma \in \Sigma_n} A_{\sigma}  \,  
         z_{\sigma(1)}^{x_1} \,  z_{\sigma(2)}^{x_2} \dots z_{\sigma(n)}^{x_n}
  \label{eq:ba} \, , 
\end{equation}
where $\Sigma_n$ is the group of the $n!$ permutations of $n$ indexes. The
coefficients $\{A_{\sigma}\}$ and the fugacities $\{z_1, \dots, z_n\}$ are
complex numbers to be determined.  The eigenvalue $E$ associated with an
eigenvector of the form~(\ref{eq:ba}) is given by
\begin{equation}
  E = -n + \sum_{i=1}^n 1/z_i \, .
  \label{eq:e}
\end{equation}
Using matching conditions at the boundary surfaces $ x_{i-1} = x_i-1$ and
the periodicity of the lattice, it can be shown that a vector $\psi$ of
the type~(\ref{eq:ba}) is an eigenvector of $M$ if the fugacities
$\{z_1,\dots, z_n\}$ satisfy the {\em Bethe equations}
\begin{equation}
 (z_i-1)^n z_i^{-L} = - \prod_{j=1}^n(1-z_j) \ \ \
   \hbox{for}    \ \ \  i=1,\dots,n  \, .
 \label{eq:be}
\end{equation} 
The procedure for deriving these equations has been thoroughly explained in
Halpin-Healy and Zhang (1995) and Derrida (1998).

The obvious solution $z_1 = \dots = z_n =1$ provides the stationary
distribution with eigenvalue $0$.  More generally, given a solution
$\{z_1,\dots, z_n\}$ of Eq.~(\ref{eq:be}), the corresponding eigenvalue
$E$ is obtained from Eq.~(\ref{eq:e}); moreover, the coefficients
$\{A_{\sigma}\}$ and thus the eigenvector $\psi$ are uniquely determined.
In order to have a complete basis of eigenvectors, $\Omega$ independent
solutions of the Bethe equations~(\ref{eq:be}) are needed.

Following Gwa and Spohn (1992), we introduce $Z_i = 2/z_i -1$.  The
equations~(\ref{eq:e}) and (\ref{eq:be}) then become, respectively,
\begin{equation}
 2E  =  -n + \sum_{j=1}^n Z_j 
  \label{eq:eZ}   \, ,  
\end{equation}
and 
\begin{equation}
  (1-Z_i)^n \ (1+Z_i)^{L-n}  
  =  - 2^L \prod_{j=1}^n \frac{Z_j - 1}{Z_j + 1}  \ \ \
  \hbox{for}    \ \ \ i=1,\dots,n      \, .
  \label{eq:bez}   
\end{equation}
We notice that the left hand side of Eq.~(\ref{eq:bez}) is a polynomial in
$Z_i$ whereas the right hand side (r.h.s.)  is independent of the index
$i$.

The analysis of the Bethe equations is simplified if only half-filled
models are considered, that is, if $L = 2n$ (Gwa and Spohn 1992).
Equation~(\ref{eq:bez}) then reduces to
\begin{equation}
  (1-Z_i^2)^n = - 4^n \prod_{j=1}^n \frac{Z_j - 1}{Z_j + 1}  \ \ \
  \hbox{for}    \ \ \ i=1,\dots,n  \, .
  \label{eq:behf}
\end{equation}
The half-filling restriction does not affect the physical behaviour of the
ASEP: in the large $L$ limit, models with arbitrary density, $\rho = n/L
\in \ ]0,1[$, belong to the same universality class. However, systems with
vanishingly small density of particles ($\rho \to 0$) or holes ($\rho \to
1$) exhibit a different behaviour and will not be discussed here.

\subsection{Analysis of Bethe equations}

Taking advantage of the fact that the r.h.s.  of Eq.~(\ref{eq:behf}) is
independent of the index $i$, the Bethe equations can be reformulated as
follows.  Consider the polynomial equation
\begin{equation}
  (1-Z^2)^n = Y \, , 
  \label{eq:z}
\end{equation}
where $Y$ is a given complex number.  Writing
\begin{equation}
 Y = -e^{u\pi} \, , 
 \label{eq:u} 
\end{equation}
$u$ being a complex number with $ -1 \le \mathrm{Im}(u) < 1$, we obtain the
$n$-th roots of $Y$
\begin{equation}
 y_m = e^{(u+i)\pi/n} e^{(m-1)2i\pi/n} \ \ \, \hbox{ for } m = 1, \ldots, n \, .
 \label{eq:ym}
\end{equation}
The $y_m$'s are evenly spaced on a circle of center 0 and radius
$|Y|^{1/n}$ and are labeled counter-clockwise $0 \le \Arg(y_1) < \Arg(y_2)
< \dots < \Arg(y_n) < 2 \pi$.  Thus, the $2n$ solutions $(Z_1, \dots,
Z_{2n})$ of Eq.~(\ref{eq:z}) are
\begin{equation}
  Z_m = (1 - y_m)^{1/2}  \,;   \ \ \ Z_{m+n} = - Z_m \ \ 
  \hbox{ with }  \,\,  m = 1, \ldots, n \, . 
  \label{eq:zm}
\end{equation}
The branch cut of the function $z^{1/2}$ is, as usual, the real semi-axis
$(-\infty,0]$, {\it i.e.}, for $m = 1, \ldots, n,$ the argument of $Z_m$
belongs to $[-\pi/2,\pi/2[$.  We explain in Appendix~\ref{ap:cassini} that
each $Z_m$ is an analytic function of $Y$ in the complex plane with a
branch cut along $[0,+\infty)$ and that the locus of the $Z_m$'s is a
remarkable curve called a Cassini oval.
 
We now choose $n$ different roots $(Z_{c(j)})_{j=1,n}$ among $(Z_1, \dots,
Z_{2n})$, such that the choice function $c: \{1, \dots, n\} \rightarrow
\{1, \dots, 2n\} $ satisfies
\begin{equation}
  1 \le c(1) < \dots < c(n) \le 2n \, .
\end{equation}
There are precisely $\Omega$ such choice functions, $\Omega = (2n)! / n!^2$
being the size of the Markov matrix.  Finally, we define
\begin{equation}
  A_c(Y) = -4^n \prod_{j=1}^n \frac{Z_{c(j)} - 1}{Z_{c(j)} + 1} \, ,
  \label{eq:ac}
\end{equation}
where $A_c$ is a function of $Y$ and of the choice function $c$.  The
Bethe equations~(\ref{eq:behf}) are equivalent to the {\it
self-consistency} equation
\begin{equation}
  A_c(Y) = Y  \, . 
  \label{eq:ay=y}
\end{equation}
Given the choice function $c$ and a root $Y$ of this equation, the
$Z_{c(j)}$'s are determined from Eq.~(\ref{eq:z}) and the corresponding
eigenvalue $E_c$ is obtained from Eq.~(\ref{eq:eZ}).

For small values of $n$, the above described procedure allows us to compute
numerical solutions of the Bethe equations. From our numerical
observations, we conjecture that for each choice function $c$ (among the
$\Omega$ possible choice functions), the self-consistency
Eq.~(\ref{eq:ay=y}) has a unique solution $Y$ that yields one eigenvector
$\psi_c$ and one eigenvalue $E_c$.  This suggests that the Bethe equations
yield a complete basis of eigenvectors for the ASEP.

Let us first consider the choice function $c(j) = j$, {\it i.e.}, the
selected $Z_j$'s are the $n$ solutions of Eq.~(\ref{eq:z}) with largest real
parts.  As this choice plays an important role in the following analysis,
we define
\begin{eqnarray}
  A_0(Y) &=& -4^n \prod_{j=1}^n \frac{Z_j - 1}{Z_j + 1} \, , \label{eq:a0} \\
  2 E_0 &=& -n + \sum_{j=1}^n Z_j  \, .\label{eq:e0}
\end{eqnarray}
We emphasize that $E_0$ is an implicit function of $Y$.  The equation
$A_0(Y) = Y$ has the solution $Y=0$ that yields $Z_j = 1$ for all $j$ and
provides the stationary distribution (or ground state) with eigenvalue $0$.

Numerical observations (Gwa and Spohn 1992) indicate that the first excited
eigenvalue $E_1$ corresponds to the choice $c(j) = j$ for $j = 1, \ldots,
n-1$ and $c(n) = n+1$; {\it i.e.}, the first excited state is obtained from
the ground state by the {\it excitation} $(n \to n+1)$ .  Writing $A_1(Y)$
and $E_1$ for the functions $A_c$ and $E_c$ corresponding to this choice
function, we have, from Eqs.~(\ref{eq:ac}, \ref{eq:a0} and \ref{eq:e0}),
\begin{eqnarray}
  A_1(Y) &=& A_0(Y) \left( \frac{Z_1-1}{Z_1+1} \ \frac{Z_n-1}{Z_n+1}
  \right)^{-1}  \, ,   
  \label{eq:a1} \\
  2E_1 &=& 2E_0 - (Z_1 + Z_n)  \, ,
  \label{eq:e1e0}
\end{eqnarray}
where we have used $Z_{n+1} = -Z_1$.  The excitation $(1 \to 2n)$, that is,
$c(j) = j+1$ for $j = 1, \ldots, n-1$ and $c(n) = 2n$, also leads to
Eqs.~(\ref{eq:a1})~and~(\ref{eq:e1e0}) and thus to the same eigenvalue
$E_1$.  The first excited state has therefore a degeneracy of order 2.

Consequently, in order to find the expression for the gap $E_1$, we must
solve the self-consistency equation
\begin{equation}
    A_1(Y) = Y  \, ,
    \label{eq:scY}
\end{equation}
calculate the $Z_j$'s for $j = 1, \ldots , n, $ and finally deduce $E_1$
from Eq.~(\ref{eq:e1e0}).

In the above discussion, we closely followed Gwa and Spohn (1992) to
present the Bethe Ansatz equations for the TASEP.  We shall now solve
these equations and calculate the gap by a radically different and simpler
method.

\section{Calculation of the gap}

Let  us define  $F(Y)$ as 
\begin{equation}
  A_0(Y) = Y \exp(F(Y)).
  \label{eq:a0f}
\end{equation}
In Appendix \ref{ap:deriv}, we have derived   the following identities, 
valid for  $|Y| \le 1$, 
\begin{eqnarray}
  F(Y)   &=& \sum_{k=1}^{\infty} \frac{w_{kn}}{k} \ Y^k \, , 
  \label{eq:fn} \\
  -4E_0 &=& \sum_{k=1}^{\infty} \frac{w_{kn-1}}{k} \ Y^k \, ,
  \label{eq:e0w}
\end{eqnarray}
where the $w_k$'s are given by
\begin{equation}
  w_k = \frac{(2k-1)!!}{(2k)!!} = \frac{(2k)!}{(k! \ 2^k)^2} \, .
  \label{eq:wk}
\end{equation}
From the Stirling formula, the leading order behaviour of $ w_k $ for $k
\to\infty$, is given by
\begin{equation}
  w_k \sim \frac{1}{\sqrt{\pi k}}  \, . 
  \label{eq:awk}
\end{equation}
From the power series~(\ref{eq:fn} and \ref{eq:e0w}) we deduce that
$A_0(Y)$ and $E_0$ are analytic functions of the complex variable $Y$
inside the unit circle. This property is not obvious {\it a priori}: the
functions $A_0(Y)$ and $E_0$, defined in Eqs.~(\ref{eq:a0}) and
(\ref{eq:e0}), respectively, depend implicitly on $Y$ via the $Z_m$'s
that involve a branch cut along $[0,+\infty)$. Indeed, for a generic
choice function $c(j)$, $A_c(Y)$ and $E_c$ are analytic only in the
complex $Y$ plane with a cut along $[0,+\infty)$ and, therefore, are not
analytic in the neighbourhood of $Y=0$.  This special property of
$A_0(Y)$ and $E_0$ is obtained by an explicit calculation in Appendix
\ref{ap:deriv} and from geometrical considerations in Appendix
\ref{ap:cassini}.

Using Eqs.~(\ref{eq:a1}),~(\ref{eq:a0f})~and~(\ref{eq:fn}), the
self-consistency equation~(\ref{eq:scY}), that determines the gap, reduces
to
\begin{equation}
 \sum_{k=1}^{\infty} \frac{w_{kn}}{k} \ Y^k 
 = \ln \frac{1-Z_1}{1+Z_1} + \ln \frac{1-Z_n}{1+Z_n} \, .
 \label{eq:sll}
\end{equation}
From  Eq.~(\ref{eq:e1e0}) and Eq.~(\ref{eq:e0w}),  we have 
\begin{equation}
  -4E_1 = \sum_{k=1}^{\infty} \frac{w_{kn-1}}{k} \ Y^k + 2Z_1 + 2Z_n \, .
  \label{eq:e1}
\end{equation}
Combining  Eq.~(\ref{eq:sll}) and  Eq.~(\ref{eq:e1}), we obtain 
\begin{equation}
  -4E_1 = \sum_{k=1}^{\infty} \frac{w_{kn}}{k(2kn-1)} \ Y^k
           + \left( 2Z_1 + \ln \frac{1-Z_1}{1+Z_1} \right)
           + \left( 2Z_n + \ln \frac{1-Z_n}{1+Z_n} \right)  \, , 
  \label{eq:e1s}
\end{equation}
 where we have used  $(2kn)w_{kn} = (2kn-1) w_{kn-1}$.

  Thus, to find the gap for a half-filled system with $n$ particles, we
  must solve Eq.~(\ref{eq:sll}) for $Y$ and substitute the result in
  Eq.~(\ref{eq:e1s}).  We emphasize that the power series in these
  equations represent, inside the unit disk, analytic functions of $Y$ that
  are defined in the whole complex plane with a cut along $[1,+\infty)$.

  We now consider the thermodynamic limit, $n \to \infty$.  We obtain, at
leading order, from Eq.~(\ref{eq:awk})
\begin{eqnarray}
  F(Y) = \sum_{k=1}^{\infty} \frac{w_{kn}}{k} \ Y^k 
     & \to & \frac{1}{\sqrt{\pi n}} \ \Li_{3/2}(Y), \\
  \sum_{k=1}^{\infty} \frac{w_{kn}}{k(2kn-1)} \ Y^k 
     & \to & \frac{1}{2\sqrt{\pi n^3}} \ \Li_{5/2}(Y),
\end{eqnarray}
where we have used the {\em polylogarithm} function of index $s$
\begin{equation}
  \Li_s(z) = \sum_{k=1}^{\infty} \frac{z^k}{k^s} \, .
\end{equation}
 By virtue of the integral representation 
\begin{equation}
  \Li_s(z) = \frac{z}{\Gamma(s)} \ \int_0^{\infty} \frac{t^{s-1} \
  dt}{e^t -z}  \, , 
  \label{eq:integr}
\end{equation}
 the function $\Li_s$ can be extended by analytic continuation to the whole
 complex plane with a branch cut along the real semi-axis $[1,+\infty)$.
 In the large $n$ limit, we deduce from Eqs.~(\ref{eq:u}, \ref{eq:ym} and
 \ref{eq:zm}) that
\begin{eqnarray}
  Z_1 = (1-y_1)^{1/2}  &=&  \sqrt{\frac{\pi}{n}} \ (-u-i)^{1/2} 
             + O \left( \frac{1}{n^{3/2}} \right)  \, , \\
  Z_n = (1-y_n)^{1/2}   &=& \sqrt{\frac{\pi}{n}} \ (-u+i)^{1/2} 
             + O \left( \frac{1}{n^{3/2}} \right)  \, ,
\end{eqnarray}
 where we  have  supposed that $ Y = -e^{u\pi}$ remains finite
 when $n \to \infty$.    Using these expressions  and the expansion 
 $ \, \ln \frac{1-Z}{1+Z} = -2Z   - \frac{2}{3} Z^3 + O(Z^5), $ 
 Eq.~(\ref{eq:sll})  reduces  to 
\begin{equation}
  \Li_{3/2}(-e^{u\pi}) = -2\pi \left[ (-u+i)^{1/2} + (-u-i)^{1/2} \right] ,
  \label{eq:li32}
\end{equation}
and  the gap~(\ref{eq:e1s}),  at  the leading order,  is given  by 
\begin{equation}
  E_1 = \frac{1}{n^{3/2}} \left\{ 
            \frac{-1}{8\sqrt{\pi}} \ \Li_{5/2}(-e^{u\pi}) 
       + \frac{\pi^{3/2}}{6} \ \left[ (-u+i)^{3/2} + (-u-i)^{3/2}
            \right] \right\}.
  \label{eq:e1li52}
\end{equation}
[Notice that the r.h.s. of Eq.~(\ref{eq:li32}) and of Eq.~(\ref{eq:e1li52})
are real when $u$ is real.]  With the help of the Maple software, we find a
unique solution of Eq.~(\ref{eq:li32}) in the strip $ -1 \le \mathrm{Im}(u)
< 1$ that is real and is given by
\begin{equation}
  u = 1.119 \, 068 \, 802 \, 804 \, 474 \dots 
\label{eq:solu}
\end{equation}
Inserting this value of $u$ in Eq.~(\ref{eq:e1li52}) yields the large $n$
(or large $L$) behaviour of the gap
\begin{equation}
  E_1 = - \frac{2.301 \, 345 \, 960 \, 455 \, 050 \dots}{n^{3/2}} 
      = - \frac{6.509 \, 189 \, 337 \, 976 \, 136 \dots}{L^{3/2}}.
\label{eq:solgap}
\end{equation}
This is precisely the result obtained by Gwa and Spohn (1992).  This gap
 scales as $L^{-3/2}$ and is real for the TASEP in the half-filling case.

\section{Summary and discussion}

  In this work, we have calculated the gap of the TASEP in the limit of a
  large size system by using the Bethe Ansatz.  We first take the large~$n$
  limit of the Bethe equations inside the unit circle, then perform the
  analytic continuation of these equations in the whole complex plane with
  a branch cut along $[1,+\infty)$ and finally solve them.  Gwa and Spohn
  (1992), on the contrary, first represent the analytic continuation of the
  Bethe equations for a fixed value of $n$ as a $n$-dependent complex
  integral (thanks to the Euler-Maclaurin formula) and then extract the
  gap from the large $n$ limit of this integral representation which is
  rather a delicate operation.  We have shown here that the derivation of the
  TASEP gap is greatly simplified by performing the calculations in the
  reverse order, that is taking the large~$n$ limit first and the analytic
  continuation afterwards.

 We do not claim that it is always true that large $n$ limit and
 analytic continuation are commuting operations. If the solution $Y$ of
 Eq.~(\ref{eq:scY}), in the large $n$ limit, diverges to $\infty$ or
 approaches asymptotically the branch cut, reversing the order of
 operations may not be possible.  Fortunately, in our problem, $Y$ remains
 a bounded negative real number when $n \to \infty$.

  We have applied our method to several other problems but, for sake of
  conciseness, we simply list these additional calculations without giving
  the details.  Our technique provides the subleading corrections to the
  gap and allows us to calculate the eigenvalue of the highest excited
  state, of a finite excitation above the ground state, or below the
  highest excited state.  The gap for the totally asymmetric exclusion
  process with an arbitrary density $\rho$ can also be obtained by
  elementary means, and we find in agreement with (Kim 1995)
\begin{equation}
  E_1\left( \rho \right)   =  
   2 \sqrt{ {\rho( 1 -\rho)} } \,\,  E_1 \left(\rho =  {1}/{2}\right)
   \pm \frac{2i\pi}{L}(2\rho -1) 
   \, . \nonumber
\end{equation}
 where $ E_1 (\rho = {1}/{2})$ is the gap for the half-filling case given
  in Eq.~(\ref{eq:solgap}); we notice that for a density other than
  one-half the gap has a non-zero imaginary part.  We have also studied
  generalizations of the ASEP by introducing a tagged particle that has the
  same dynamics as the other particles: the gap then scales as $L^{-5/2}$.

  Finally, we emphasize that the formula~(\ref{eq:integr}) already appeared
  in the work of Derrida and Appert (1999): indeed, Eqs.~(\ref{eq:sll}) and
  (\ref{eq:e1}) are similar to those used in their calculation of large
  deviation functions of the ASEP by Bethe Ansatz.

\subsection*{Acknowledgments}

It is a pleasure to thank C. Appert, B. Derrida, C. Godr\`eche, J.-M. Luck
and S. Mallick for inspiring discussions and remarks about the manuscript.

\appendix 

\section{Derivation of Eq.~(\ref{eq:fn}) and Eq.~(\ref{eq:e0w})}
\label{ap:deriv}

The numbers $w_k$ defined in Eq.~(\ref{eq:wk}) are the coefficients of the
  Taylor series
\begin{equation}
   \frac{1}{\sqrt{1-x}} = \sum_{k=0}^{\infty} w_k x^k \, . 
\end{equation}
By integration, we find  
\begin{equation}
  \sqrt{1-x} = 1 - \frac{1}{2} \ \sum_{k=1}^{\infty} \frac{w_{k-1}}{k} x^k 
  \, . 
\end{equation}
 Recalling that $Z_m = \sqrt{1-y_m}$ [see
 Eqs.~(\ref{eq:ym})~and~(\ref{eq:zm})], we obtain
\begin{equation}
  \sum_{m=1}^n Z_m = n - \frac{1}{2} \ \sum_{k=1}^{\infty}
             \frac{w_{k-1}}{k} \sum_{m=1}^n  y_m^k  \, . 
  \label{eq:sumZ} 
\end{equation}
 The fact that  the  $y_m$'s  are the  $n$-th roots  of $Y$ leads to the
 following  relation 
\begin{equation}
   \sum_{m=1}^n y_m^k = \left\{ \begin{array}{ll}
       nY^{k/n} & \mbox{if $k$ is a multiple of $n$} \\
       0        & \mbox{otherwise.}     
                               \end{array}      \right.
   \label{eq:ymk}
\end{equation}
  Inserting this relation   in  Eq.~(\ref{eq:sumZ}),  we obtain 
\begin{equation}
  \sum_{m=1}^n Z_m = n - \frac{1}{2} \ \sum_{k=1}^{\infty}
             \frac{w_{nk-1}}{k} Y^k  \, .  
  \label{eq:sumZinY}
\end{equation}
 We thus find, thanks to the crucial identity~(\ref{eq:ymk}), that
 $\sum_{m=1}^n Z_m$ is analytic in $Y$ inside the unit circle.  Finally,
 substituting Eq.~(\ref{eq:sumZinY}) in Eq.~(\ref{eq:e0}), we obtain
 Eq.~(\ref{eq:e0w}).

The derivation of Eq.~(\ref{eq:fn}) follows similar steps.  We first notice
 that the Taylor expansion of the function
\begin{equation}
  f(x) = \ln\left( \frac{4}{x}\ \frac{1-\sqrt{1-x}}{1+\sqrt{1-x}} \right)
  \,   \label{eq:deff}
\end{equation}
 is  given by 
\begin{equation}
  f(x) = \sum_{k=1}^{\infty} \frac{w_k}{k} x^k \, .  \label{eq:taylorf} 
\end{equation}
 [This follows from   $f(0) = 0$ and 
 $  f'(x) = \frac{1}{x} \left(\frac{1}{\sqrt{1-x}} - 1 \right)\,].$
 Using   Eq.~(\ref{eq:a0}) and  the identity $\prod_{m=1}^n (-y_m) = -Y, $
 we deduce that 
\begin{equation}
   A_0(Y) = Y \prod_{m=1}^n \frac{4}{y_m}\ \frac{1-Z_m}{1+Z_m}  \, .
 \label{eq:prodA} 
\end{equation}
 From Eq.~(\ref{eq:a0f}) and Eqs.~(\ref{eq:deff}),~(\ref{eq:taylorf}) and
  ~(\ref{eq:prodA}), we obtain
\begin{equation}
  F(Y) =  \sum_{m=1}^n f(y_m) = \sum_{k=1}^{\infty} \frac{w_k}{k} \sum_{m=1}^n
  y_m^k \, .
\end{equation}
  This equation reduces to Eq.~(\ref{eq:fn}) by virtue of the
  identity~(\ref{eq:ymk}).

\section{Roots  of the Bethe Equations  and Cassini ovals}

\label{ap:cassini}

 The polynomial equation
\begin{equation}
  (1-Z^2)^n = Y \,
\end{equation}
where $Y$ is a fixed complex number, has $2n$ solutions, $(Z_1, \dots,
Z_{2n})$.  The purpose of this appendix is to explain how these solutions
can be labeled in a coherent way so that each root $Z_m(Y)$ is an analytic
function of $Y$.

 We first notice that $y_m$, defined in Eq.~(\ref{eq:ym}), is an analytic
function of $Y$ in the complex plane with a branch cut along the real
semi-axis $[0,+\infty)$.  Nevertheless when $Y$ crosses $[0,+\infty)$, the
functions $y_1, y_2 \dots, y_n$ are the analytic continuations (above the
axis) of respectively $y_n, y_1, \dots, y_{n-1}$: thus the existence
of the branch cut along $[0,+\infty)$ is due to the labeling of the roots.

\begin{figure}
  \centering
  \includegraphics[width=\figwidth, keepaspectratio]{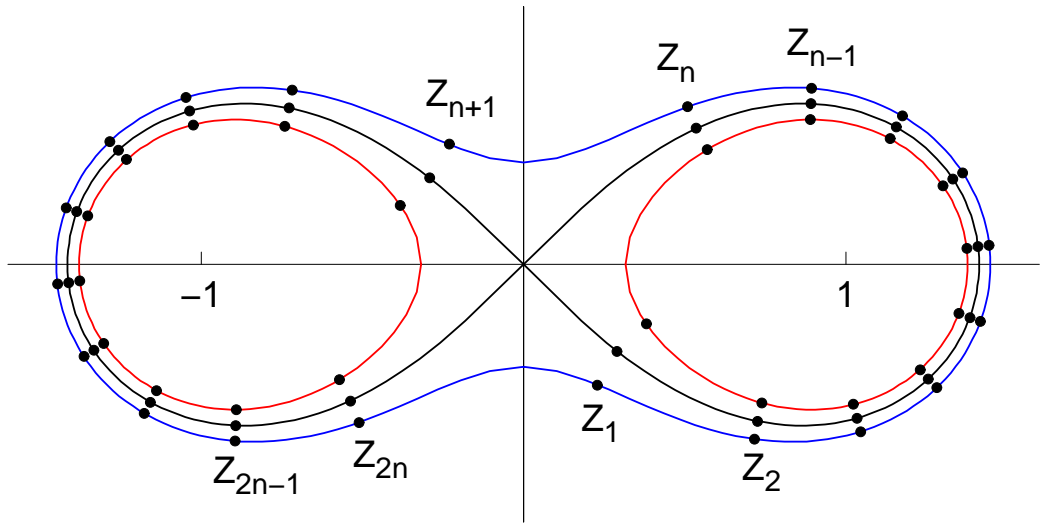}
  \caption{\em 
     Labeling the roots $Z_m$ of the equation $(1-Z^2)^n=Y$.
     Here $Y = e^{i\phi} r^n$ with $n=10$, $\phi = \pi/2$, and $r \in
     \{0.9, 1, 1.1\}$.  The continuous curves are the corresponding Cassini
     ovals (see text).  When $r$ is fixed and $\phi$ goes from 0 to $2\pi$,
     each $Z_m$ slips counter-clockwise along the Cassini ovals.  Then, the
     jump $\phi = 2\pi \to 0$, {\rm i.e.}, $Y$ crosses $[0,+\infty)$,
     consists of a global shift of the labels $m$ around each continuous
     curve.
  }
  \label{fig}
\end{figure}

 The complex numbers $(Z_1, \dots, Z_{2n})$ belong to the curve defined by
\begin{equation}
  \left|Z-1 \right|  \,   \left| Z+1\right| = r   \,\,\, 
  \hbox{ with  } \,\,\, r = |Y|^{1/n}   \, 
\label{eq:cassini}
\end{equation}
  and called a {\em Cassini oval}.  A Cassini oval is the conformal
  transformation of the circle of center 1 and radius $r$ by the function
  $z \rightarrow z^{1/2}$.  Its shape depends on whether the point 0 is
  inside or outside the circle, {\it i.e.}, whether $r<1$ or not.  When
  $r<1$, the curve of equation~(\ref{eq:cassini}) consists of two ovals
  around the points $Z=\pm 1$.  The numbers $(Z_1,\dots,Z_n)$ lie on the
  right oval and $(Z_{n+1},\dots, Z_{2n})$ on the left oval.  For the
  marginal case, $r=1$, the curve is the {\em lemniscate of Bernoulli},
  with a multiple point at $Z=0$.  When $r>1$, the Cassini oval is a single
  loop with a peanut shape (when $r \in ]1,2[$) or an oval shape ($r \ge
  2$).  See Fig.~\ref{fig} where the cases $r = 0.9, 1$ and $1.1$ are
  drawn.  In the large-$r$ limit, the oval tends to the circle of radius
  $\sqrt{r}$.

 We now discuss the analyticity properties of $Z_m(Y)$: $Z_m$ is an
 analytic function of $y_m$ with a branch cut along $(-\infty,0]$; this
 branch cut is compatible with that of $y_m(Y)$.  Consequently $Z_m$ is
 an analytic function of $Y$ with a branch cut along $[0,+\infty)$.
 Moreover when $Y$ crosses the real segment $[0,1[$, the functions $Z_1(Y),
 Z_2(Y) \dots, Z_n(Y)$ are the analytic continuations (above the axis) of
 the functions $Z_n(Y), Z_1(Y), \dots, Z_{n-1}(Y)$ respectively.  See
 Fig.~\ref{fig}.  (A similar property is true for the functions
 $Z_{n+1}(Y), Z_{n+2}(Y) \dots, Z_{2n}(Y)$ that belong to the left oval).
 But when $Y$ crosses $]1,+\infty)$, the functions $Z_1(Y), Z_2(Y), \dots,
 Z_{2n}(Y)$ are the analytic continuations (above the axis) of the
 functions $Z_{2n}(Y), Z_1(Y), \dots, Z_{2n-1}(Y)$ respectively.
 Consequently the branch cut of the function $A_0(Y)$ defined in
 Eq.~(\ref{eq:a0}) is $[1, +\infty)$ and not $[0, +\infty)$.

\section*{References}
\begin{itemize}

\item 
F.~C.~Alcaraz, M.~Droz, M.~Henkel, V. Rittenberg,  1994,
{\em Reaction-diffusion processes, critical dynamics and quantum
 chains},
 Ann. Phys.  {\bf 230}, 250.

\item 
F.~C.~Alcaraz and M.~J.~Lazo, 2003,
{\em The Bethe Ansatz as a matrix product Ansatz}, 
 preprint  cond-mat/0304170. 

\item 
 R. Bundschuh, 2002, 
{\em Asymmetric exclusion process and extremal statistics of
 random sequences},
Phys. Rev. E {\bf 65}, 031911.

\item 
 B. Derrida, 1998, 
{\em An exactly soluble non-equilibrium system: the asymmetric simple
 exclusion process},  
 Phys. Rep.  {\bf 301}, 65.

\item 
  B. Derrida, C. Appert, 1999,
{\em Universal large-deviation function of the Kardar-Parisi-Zhang
 equation in one dimension},  
 J. Stat. Phys. {\bf 94}, 1. 

\item 
 B. Derrida, M. R. Evans, 1999,
 {\em Bethe Ansatz solution for  a defect particle in 
the asymmetric exclusion process}, 
 J. Phys. A: Math. Gen. {\bf 32}, 4833.

\item 
 B. Derrida, M.~R.~Evans, V. Hakim, V. Pasquier, 1993,
 {\em Exact solution of a 1D  asymmetric exclusion
model  using a matrix formulation}, 
J. Phys. A: Math. Gen. {\bf 26}, 1493.

\item 
 B. Derrida, J. L. Lebowitz, 1998,
{\em Exact  large deviation function in the asymmetric exclusion process},  
 Phys. Rev. Lett.  {\bf 80}, 209.

\item 
 B.~Derrida, J.~L.~Lebowitz, E.~R.~Speer, 2003,
{\em Exact  large deviation functional  of a 
 stationary open driven diffusive system:  the asymmetric exclusion process},  
 J. Stat. Phys. {\bf 110},  775.

\item 
  D. Dhar, 1987,
{\em    An exactly solved model for interfacial growth},
  Phase Transitions {\bf 9}, 51. 

\item 
 M.~R.~Evans, D.~P.~Foster, C.~Godr\`eche, D.~Mukamel, 1995,
 {\em  Asymmetric exclusion model with two species: spontaneous
 symmetry breaking}, 
 J. Stat. Phys. {\bf 80},  69.

\item 
 M.~R.~Evans, Y.~Kafri, H.~M.~Koduvely,  D.~Mukamel, 1998,
{\em  Phase separation in one-dimensional driven diffusive systems},
  Phys. Rev. Lett.  {\bf 80}, 425.

\item 
 L.-H. Gwa, H. Spohn, 1992,
  {\em Bethe solution for the dynamical-scaling exponent of the noisy
  Burgers equation},
  Phys. Rev. A {\bf 46}, 844.

\item 
 T. Halpin-Healy, Y.-C.~Zhang, 1995, 
{\em Kinetic roughening phenomena, stochastic growth, directed polymers and
 all that},
Phys. Rep.  {\bf 254}, 215.

\item 
  D. Kim, 1995,
 {\em Bethe Ansatz solution  for crossover scaling functions
 of the asymmetric XXZ chain  and the Kardar-Parisi-Zhang-type
 growth model},
 Phys. Rev. E {\bf 52}, 3512.

\item 
 J. Krug, 1991,
 {\em Boundary-induced phase transitions  in  driven diffusive systems},
Phys. Rev. Lett.  {\bf 67}, 1882.

\item 
J. Krug, 1997,
 {\em Origins of scale invariance in growth processes}, 
 Adv.   Phys. {\bf 46}, 139.

\item 
D. G. Levitt, 1973,
 {\em Dynamics of a single-file pore: Non-Fickian behavior}, 
 Phys. Rev. A {\bf 8}, 3050.

\item 
 C.~T.~MacDonald, J.~H.~Gibbs, 1969,
  {\em Concerning the kinetics of poly\-peptide synthesis
 on polyribosomes}, 
Biopolymers {\bf 7}, 707. 

\item 
 P. M. Richards, 1977,
 {\em Theory of one-dimensional hopping conductivity and diffusion}, 
 Phys. Rev. B {\bf 16}, 1393.

\item 
 M.~Schreckenberg, D.~E.~Wolf (ed.), 1998,  
  {\em  Traffic and granular flow '97}
 (Berlin: Springer-Verlag).

\item 
 G.~Sch\"utz, 1993,
 {\em Generalized  Bethe Ansatz solution of a one-dimen\-sional
  asymmetric exclusion  process on a ring with blockage},
 J. Stat. Phys. {\bf 71},  471.

\item 
E.~R. ~Speer, 1993,
{\em The two species totally  asymmetric exclusion  process},
 in Micro, Meso and Macroscopic approaches in Physics, M.~Fannes
 C. Maes and  A. Verbeure Ed. NATO Workshop 'On three levels',
 Leuven, July 1993.

\item 
H. Spohn, 1991,
{\em Large scale dynamics of interacting particles},
 (New-York: Springer-Verlag).

\item 
 R.~B.~Stinchcombe and G.~M.~Sch\"utz, 1995,
 {\em  Application of operator algebras  to stochastic dynamics
  and the Heisenberg chain}, 
 Phys. Rev. Lett.  {\bf 75}, 140.

\item 
B.~Widom, J.~L.~Viovy, A.~D.~Defontaines, 1991,
{\em  Repton model of gel electrophoresis and diffusion},
J. Phys. I France {\bf 1},  1759.

\end{itemize}

\end{document}